\documentclass[12pt]{article}
\usepackage{epsf}
\usepackage{graphicx}
\usepackage{amsmath}
\title{ On the Shape Dependence of the Tangential Casimir Force }
\author{
Yu.S.Voronina,\footnote{E-mail: voronina-yulya@yandex.ru} \ \ 
P.K.Silaev
\\ \it Moscow State University, Physics Dept. \\
\small PACS numbers: 03.70.+k, 12.20.Ds 
}
\begin{document}
\maketitle
\begin{abstract}
The normal and tangential Casimir force for the rack 
gear is calculated numerically in the case of ideal boundary conditions for the electromagnetic field --- perfect reflection on the boundaries. The resulting tangential force appears to be essentially shape dependent. Relatively small shape variations lead to the essential changes in tangential force, whereas normal force remains almost unchanged. 
\end{abstract}
\section{Introduction}
During last two decades a large number of studies, both theoretical and experimental, deals with the Casimir effect \cite{casimir}. 
The growing interest to this problem is motivated by experimental results \cite{exp1,exp2,ref1,ref2}, that provide relatively precise 
confirmation of QFT predictions that lays out of the bounds of particle physics. On the other hand, Casimir effect leads to the possibility of friction-free nanomechanical devices.
Recent studies have aimed to find configuration with repulsive Casimir force without dielectric fluid. 
So the Casimir force was calculated for different geometry configurations, especially
for the configurations with repulsive and tangential forces. 
Some configurations are smooth (sphere, cylinder, parabolic cylinder, plate \cite{sphere,cylinder,knife}), whereas some configurations are sharp (wedge, cone, knife, needle \cite{knife,knife2,knife3,sharp}) or simply rectangle (flat metallic surfaces with $\pi/2$ angles between them). 

It is well-known, that in the classical electrodynamics the precise shape of bodies (sharp edges, needles, etc) have a significant influence on the corresponding electromagnetic solutions.
So the question arises, whether the small changes of shape can change the results for Casimir 
force? From \cite{sharp} one can conclude that it is quite possible. For instance, even for
a cone with finite angle we observe additional singularity, caused by the vertex of the cone. 
It should be stressed, that in \cite{knife} the case of sharp knife edge is considered as a limiting case 
of parabolic cylinder, with the smooth dependence on the parameter of parabolic curve. Our approach is quite different: we preserve the "global" shape properties, and change the geometry only in the nearest vicinity of the edges.

\begin{figure}
\includegraphics{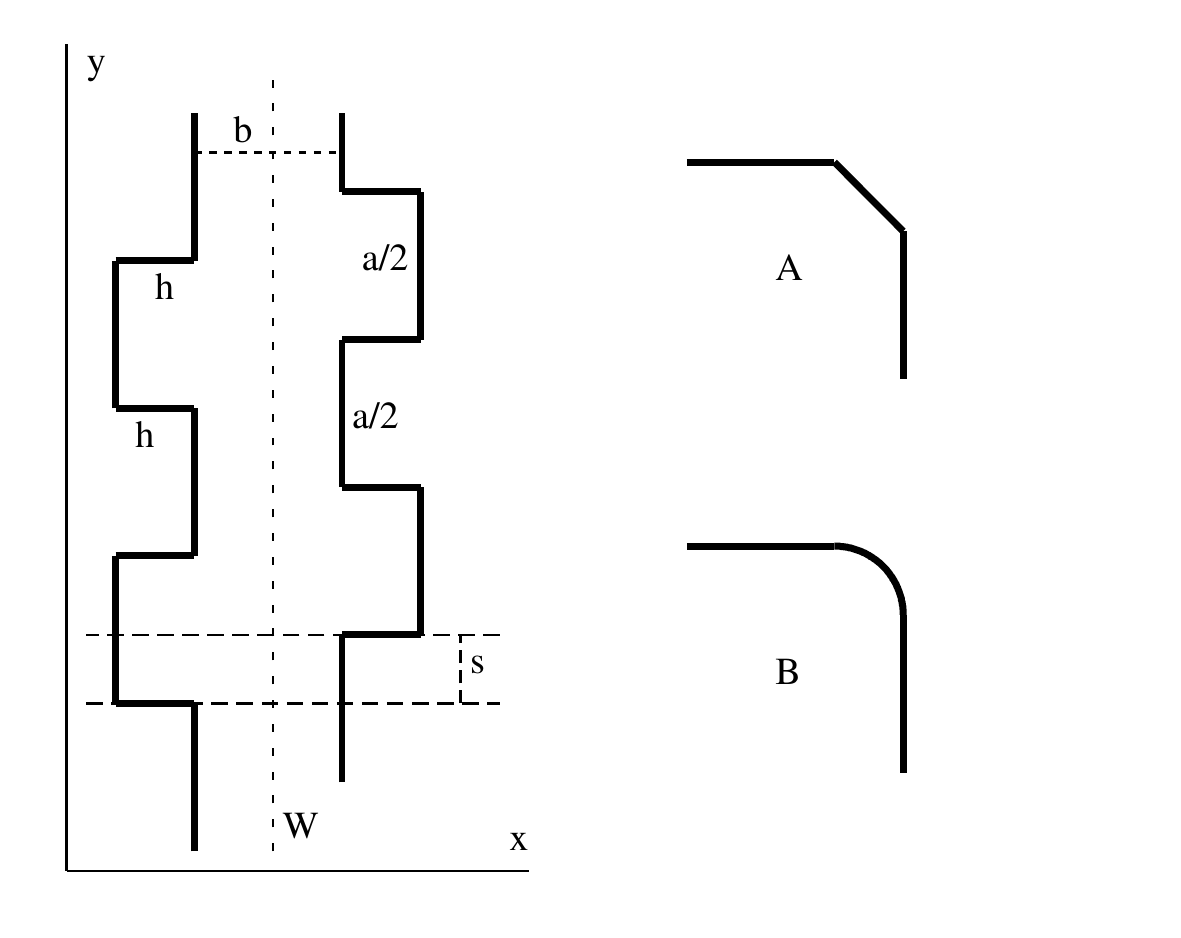}
\caption{System parameters and shape of edges: a --- period, h --- depth of profile, s --- shift, b --- width of the gap. Dotted line W denotes the surface in the gap between profiled plates. Case A: flat edge. Case B: smooth edge. }
\end{figure}

Let's consider the rack gear (see Fig. 1. for geometry details). 
We have two profiled plates, parallel to y axis, period of profiles is $a$, distance between plates is $b$, and the relative shift of plates is $s$. This geometry is translational invariant along z axis (z axis is orthogonal to picture plane). For simplicity we consider the ideal 
case: material of plates is perfect metal (perfect boundary reflection), without any realistic frequency dependence. We also will ignore the temperature dependence.
So we consider the following question: what happens, if we change all
$\pi/2$ angles of this rectangular geometry by the edges? We will investigate two cases: 
case A --- flat edge, where the $\pi/2$ angle is replaced by two $3\pi/4$ angles, 
and case B --- smooth edge, where $\pi/2$ angle is replaced by cylinder of appropriate radius (see Fig. 1).

From the standard explanation of the tangential Casimir force for this geometry one can conclude, that 
it shouldn't be any dependence on the edge shape.
Indeed, for fixed gap width $b$ in the case of zero shift ($s=0$) the average distance between our bodies 
(plates) is less, then in the case of half-period shift ($s=a/2$). 
So we should obtain the tangential force, that cannot depend on the shape details. 
However, this conclusion is correct only if we can neglect side effects at the
edges of profile wells, but for the system considered the side effects are significant (see sec. 2). Side effects can essentially depend on the shape details, so 
it is possible to observe shape dependence of the tangential Casimir force for
our geometry.

\section{Calculation method}

One of the most efficient ways to estimate force between two isolated bodies is to calculate energy-momentum tensor for the vacuum state on the 
surface $W$ enclosing one of the bodies
$$F_i=\oint\limits_{W}\langle 0|T_{ij}|0\rangle dS_j.$$
Vacuum expectation value of the energy-momentum tensor can be  expressed in terms of euclidian Green function via integral over pure imaginary frequency $\omega(\mu)=i\mu$.
Originally such approach was suggested in \cite{Dzyaloshinskii} for evaluation van der Waals forces between two bodies. Recently this method has been advanced for numerical calculation of the Casimir force \cite{Rodriguez1,Rodriguez2}. 
It is interesting to note that the selection of the integration contour $\omega(\mu)=i\mu$ is not unique.
The employment of different variant of the frequency contour were considered in \cite{Rodriguez2}.
Here the problem is formulated at real "frequency" $\mu$ and effective complex dielectric permittivity.

For the problem in question the surface $W$ separating two plates can be chosen as y-z plane (see Fig. 1). After integration on y 
\begin{equation}
F_n=\int_{y_0}^{y_0+a} \langle 0|T_{11}|0\rangle\;dy, \qquad F_\tau=\int_{y_0}^{y_0+a} \langle 0|T_{12}|0\rangle \;dy
\end {equation}
for arbitrary $y_0$
we obtain the normal and tangential force, corresponding to one period $a$ in $y$-direction and unit length in $z$-direction.
Then we divide these values (for one period) by the period length, and so we get the "density" of both forces.

For further computation euclidian Green function $G_{ij}(\mu,u,v)$ should be constructed, that is 
transversal:

\begin{equation}
\partial_{u_i}G_{ij}(\mu,\vec u,\vec v)=0, \qquad \partial_{v_j}G_{ij}(\mu,\vec u,\vec v)=0, 
\end {equation}
it also satisfy boundary conditions 
\begin{equation}
\tau_i G_{ij}(\mu,\vec s,\vec v)=0=\tau_j G_{ij}(\mu,\vec u,\vec s), 
\end {equation}
where $\vec \tau$ is tangent vector to the boundary surface at a point $\vec s$,
and, finally, it should be solution to equation 
\begin{equation}
\triangle G_{ij}(\mu,\vec u,\vec v) - \mu^2 G_{ij}(\mu,\vec u,\vec v) = \delta_{ij}(\vec u-\vec v),
\end {equation}
where 
$$\delta_{ij}(\vec u-\vec v)=\int\limits_{-\infty}^{\infty}d\vec k \left(\delta_{ij}-{k_i k_j\over k^2}\right) {e^{i\vec k (\vec u-\vec v)}\over(2\pi)^3}$$ 
is transversal delta function. 
One can easily verify, that 
\begin{equation}
\langle 0|A_i(\vec u)A_j(\vec v)|0\rangle=-{1 \over \pi}\int_0^\infty d\mu \; G_{ij}(\mu,\vec u,\vec v). \label{Vev_A}
\end {equation}
Energy-momentum tensor is constructed of the derivatives of the left hand side on $u_\alpha$ and $v_\beta$ for $\vec v=\vec u$, so we can 
estimate force, provided we solve equations for euclidian Green function.
Now we should take into account the translational invariance on $z$.
We perform Fourier transformation along $z$ axis. 

Definitely, one can also construct the series along $y$ axis, because our geometry is periodic in y-direction, but straightforward solution 
of two-dimensional equation 
appears to be more simple and effective. 
After Fourier transformation along $z$ axis we get 
\begin{equation}
G_{ij}(\mu,\vec u,\vec v)={1\over 2\pi}\int dq \exp(iq(u_3-v_3))G_{ij}(q,\mu,\vec u_2,\vec v_2)
\end {equation}
where $\vec u_2$ and $\vec v_2$ are two-dimensional vectors, constructed from the first two components of $\vec u$ and $\vec v$ respectively. 
Finally, for the arbitrary term in the expression of energy-momentum tensor for vacuum state we obtain
\begin{equation}
\left.\langle 0|\partial_{u_\alpha}A_i(\vec u)\partial_{v_\beta}A_j(\vec v)|0\rangle\right|_{\vec u=\vec v}=-\left.{1 \over 2\pi^2}\int dq\, \partial_{u_\alpha}\partial_{v_\beta}\int_0^\infty d\mu \; G_{ij}(q,\mu,\vec u_2,\vec v_2)\right|_{\vec u_2=\vec v_2}. \label{Vev_dA}
\end {equation}
Here $\partial_{u_3}\equiv iq$ and $\partial_{v_3}\equiv -iq$. For distant bodies renormalization of this expression is trivial --- 
subtraction of Green function $ G_{ij}^0(q,\mu,\vec u_2,\vec v_2) $ for Minkowski space yields finite expression.

Therefore, renormalized vacuum expectation values (\ref{Vev_A}), (\ref{Vev_dA}) are expressed through the difference $ G^{ren}_{ij}(q,\mu,\vec u_2,\vec v_2) = G_{ij}(q,\mu,\vec u_2,\vec v_2)- G_{ij}^0(q,\mu,\vec u_2,\vec v_2) $, which is the solution of the homogeneous Helmholtz equation with the boundary condition $\tau_j G^{ren}_{ij}(q,\mu,\vec u_2,\vec s_2)=-\tau_j G^{0}_{ij}(q,\mu,\vec u_2,\vec s_2)$, where $\vec s_2$ 
lays on the plate surface.
One of the most effective approach to solving the problem for $G^{ren}_{ij}(q,\mu,\vec u_2,\vec v_2)$ is the boundary-element method (BEM) \cite{bem1,bem2,bem3}. In the present work we apply slightly modified BEM: instead of standard spline approach for the given boundary element we use polynomial approximation, based on surrounding elements. 
It is similar, but not precisely equal to spline.
Additionally, we impose more conditions, then the number of elements and just minimize discrepancy in resulting overdetermined linear system.
Both modifications increase precision for given number of elements.

To estimate errors of this computation scheme, we perform calculations for the trivial case $h=0$, with well-known explicit analytical result. 
In further calculations we use the number of point per unit length of the boundary that in this trivial case yields relative error about $10^{-4}$. 
We also estimate error for our nontrivial boundary by increasing points density and subsequent comparison of results. 
It appears, that nontrivial boundary form increase relative error up to $10^{-3}$.

\begin{figure}
\includegraphics{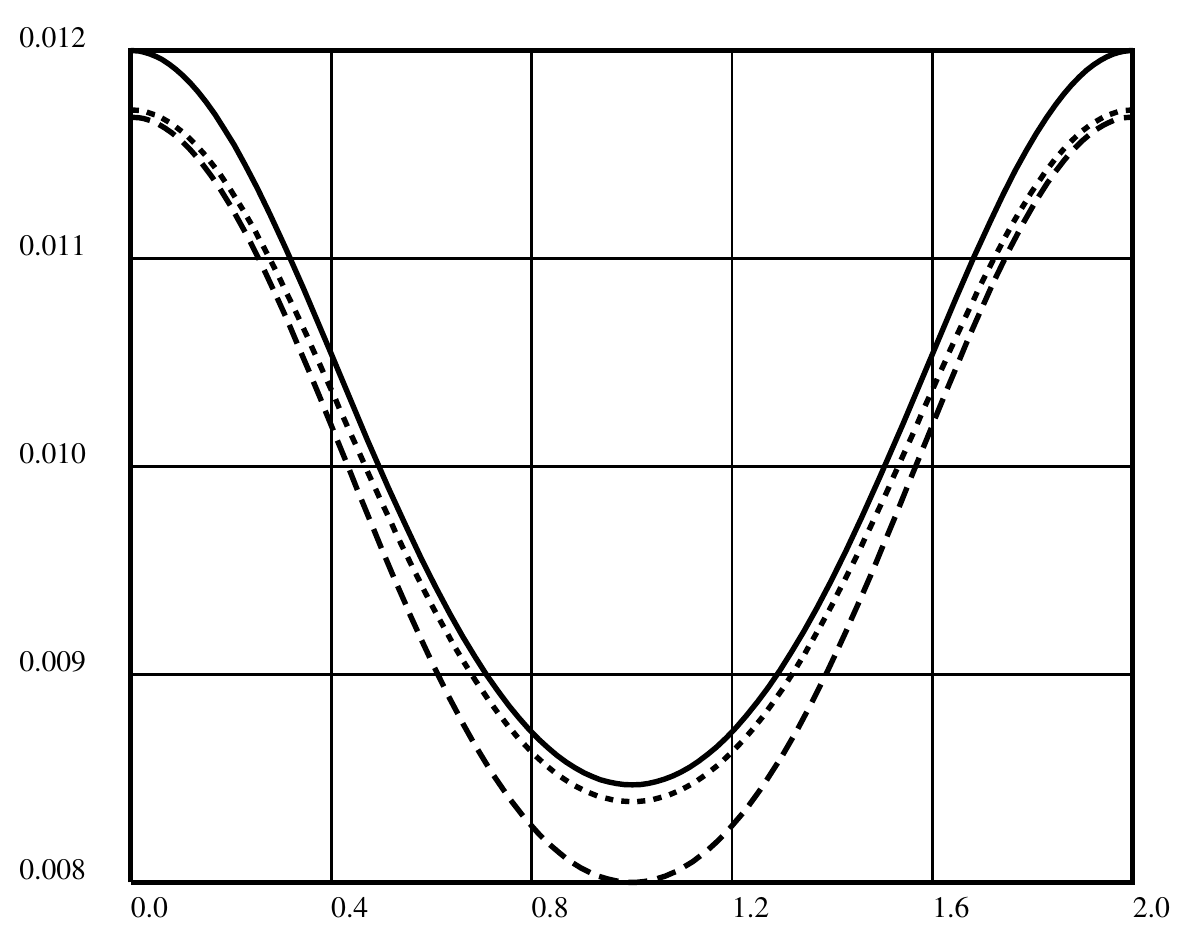}
\caption{Density of normal force as the function of shift s. Rectangle case --- solid line, flat edge --- dashed line, smooth edge --- dotted line. }
\end{figure}

\section{Results and discussion}

We use natural system of units $\hbar=c=1$ and choose the geometry parameters $a=2$, $h=0.5$, $b=1$. 
This choice provide us comparable order of magnitude for tangential and normal force density. For the trivial case $h=0$ the density of normal force will be equal to $\pi^2/240=0.0411$, and this value will be used as a reference point.
For plates with profile $A$ the edge lengh is $\sqrt{2}L=\sqrt{2}\cdot 0.08$ and in case $B$ radius of the corresponding cylinder is $R=0.08$. 

\begin{figure}
\includegraphics{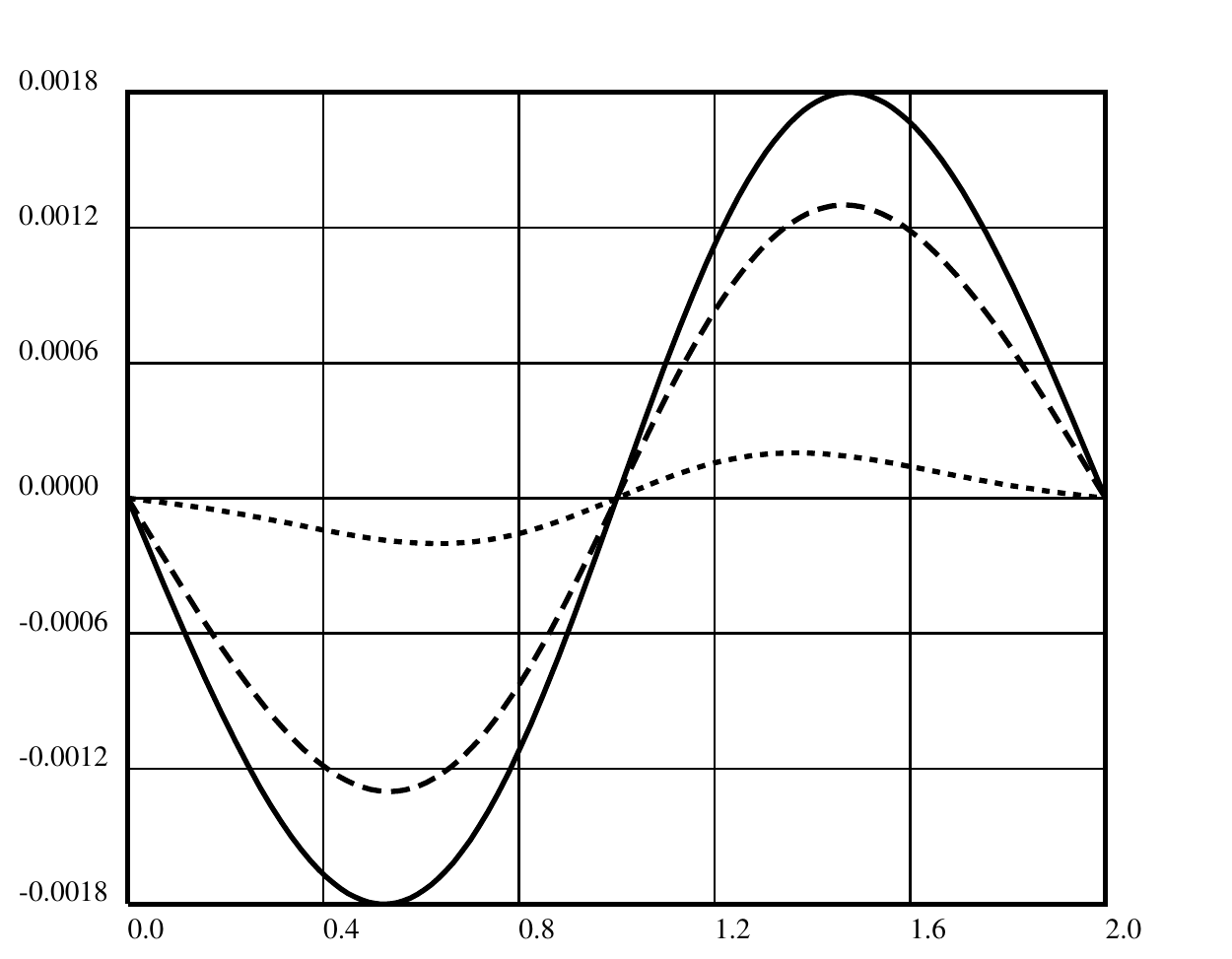}
\caption{Density of tangential force as the function of shift s. Rectangle case - solid line, flat edge --- dashed line, smooth edge - dotted line. }
\end{figure}

From Fig. 2 one can easily find, that value of normal force density $f_n$ (it should be stressed, that here "density" means value, calculated for one period, divided by the length of period) for
zero shift $s$=0 ($f_n$=0.0124) essentially differs from the half of our reference value (0.0411/2=0.0205). 
So side effects are significant. At the same time the shape effects for zero shift can be neglected, because they are even smaller, then trivial estimation of these effects calculated with the proximity force approximation (PFA).
Let us notice that we do not use the PFA directly because of its low accuracy. 
Instead we multiply the normal force density for plates with edge calculated with the PFA by the normalization constant defined as the ratio of forces for rectangular geometry $N={f_n \over f^{PFA}_n}$, where $f^{PFA}_n={\pi^2\over 480 a}\left({1\over b^4}+{1\over (b+2h)^4}\right)$ and the value $f_n$ was calculated with the use of the method described in previous section.
For both flat and smooth edge we get approximately $f_n$=0.0121, whereas PFA estimation yields $N\cdot f_n^{PFA}\approx 0.0115$ in case A, where
\begin{multline}
f_n^{PFA}={\pi^2\over 720 a^2}\left( {1\over (b+2h-2L)^3} -{1\over (b+2h)^3} + \right.\nonumber\\
\left. + 3\,{{a\over 2} - 2L \over b^4}\left(1+{b^4\over (b+2h)^4}\right) + {2L\over b^2 (b+2L)^2}\left(4\,{(b+L)^2\over b(b+2L)}-1\right) \right),
\end{multline}
and $N\cdot f_n^{PFA}\approx 0.0111$ in case B, where 
\begin{multline}
f_n^{PFA}={\pi^2\over 240 a^2}\left( {\left({a\over 2}-2R\right)\over b^4 } + 2\int\limits_0^R {dy \over \left( b + 2 \left(R-\sqrt{R^2-(R-y)^2}\right) \right)^4} +   \right.\nonumber\\
\left.  + {\left({a\over 2}-2R\right)\over (b+2h)^4 } +  2\int\limits_0^R {dy \over \left( b + 2h - 2 \left(R-\sqrt{R^2-(R-y)^2}\right) \right)^4} \right).
\end{multline}

For the half-period shift $s$=1 we obtain almost identical behavior. Side effects are
significant, whereas shape effects are relatively small. The difference between flat edge on the one side, and smooth edge and rectangle case on the other side can be explained, if
we take into account that average distance between bodies is definitely
greater for the flat edge (in the case of half-period shift). 
So we can conclude, that for normal force shape effects can be neglected, 
the change in force density appears to be about 2-3\%.

On the contrary, for the density of tangential force we obtain essential shape dependence (see Fig. 3). Even absolute value of the difference of tangential forces between rectangle case, flat edge case and smooth edge case
is much greater, then for normal force. Let us remind, that for the flat edge we just change
one $\pi/2$ angle to two $3\pi/4$ angles, whereas for the smooth edge we have no angles at all.
So the dependence on the shape of edge seems to be quite reasonable.

It should be noted, that different dependence on the edge shape for normal and tangential force is quite reconcilable with energy reasons. Different energy functions for different shapes can have almost identical derivatives on the variable $x$ (normal force), 
and quite different derivatives on the variable $y$ (tangential force). Even for normal force 
we observe different dependence on the shift $s$ for different edge shapes (see Fig. 2).

For both normal and tangential force we obtain the regular dependence on the edge size.
For example, in the case of flat edge and shift $s$=0.6 we get:

\vskip 4 mm
\begin{tabular}{|p{3 cm}|c|c|c|c|}\hline
\strut{\small Edge size} & 0 & 0.04 & 0.08 & 0.12 \\ \hline
\strut{\small Density of tangential force} & 0.00178 & 0.00153 & 0.00129 & 0.00110 \\ \hline
\end {tabular} 
\vskip 4 mm
 
For other values of shift $s$ we observe the same type of dependence. And for the case of smooth edge we also obtain this regular dependence. 

\section{Conclusion}

Direct numerical computations lead us to the conclusion, that (at least for the geometry considered) there is essential dependence of the Casimir force on the details of geometry.
If we change rectangle structures by the structure with edges (flat or smooth), it leads to essential variation of the 
tangential force. 
It should be noted, that variation of the force value isn't proportional to the relative 
size of edges: maximum edge size we used was 0.14 --- about 15\% of typical length in our geometry, whereas tangential force for smooth edge appears to be 8 times smaller, then in
rectangular case. 

The result observed is definitely pure side effect, but sometimes side effects play significant role. 
It also should be mentioned, that we consider the case of perfect metallic surface (perfect mirror). 
The influence of realistic frequency dependence on the shape effects is not obvious. 
Probably, it can mask all these effects, but this can be specified only by the direct calculations. 
As for temperature dependence, from \cite{sharp} we can conclude, that non-zero
temperature can only amplify the effect observed, because for flat surface and sharp edges the temperature dependence appears to be quite different. 

\section{Acknowledgements}
The authors are grateful to Prof. K.Sveshnikov for interest and support, and to Dr. O.Pavlovski and M.Ulybyshev for valuable discussions.

\end{document}